# Safe Farming: Development of a Prevention System to Mitigate Vertebrates Crop Raiding

Razi Iqbal
E-mail: razi.iqbal@ieee.org

One of the main problems for farmers is the protection of their crops, before and after harvesting, from animals and birds. To overcome this problem, this paper proposes a model of safe farming in which the crops will be protected from vertebrates' attack through a prevention system that is based on Wirelesses Sensors Networks. Different sensor nodes are placed around the field that detect animals or birds' existence and generate required signals and information. This information is passed to the Repelling and Notifying System (RNS) that is installed at the field through a short range wireless technology, ZigBee. As RNS receives the information, it generates ultrasonic sounds that are unbearable for animals and birds, which causes them to run away from the field. These ultrasonic sounds are generated in a frequency range that only animals and birds can hear, while humans cannot notice the sound. The paper also proposes a notifying system. It will inform the farmer about animals or birds' intrusion in the field through SMS, but doesn't need any action from the farmer. The low cost and power efficiency of the proposed system is a key advantage for developing countries where cost and power are major players in any system feasibility.

## 1. INTRODUCTION

The main objective of this research is to develop a prevention system for the famers of developing countries that will help them prevent their farms' crops (field) from vertebrates (animals and birds with a backbone). Farmers use different methods to protect their crops from vertebrates. For a long time, farmers have used scarecrows (a man statue made up of dry grass) in their fields, but that doesn't seem to work. This results in damage of crops with zero prevention. Staying in the farms 24/7 and guarding the crops is not possible because of different reasons e.g. weather, security concerns and sometimes money, because not all the farmers can afford hiring a person who can guard their fields in their absence. These all problems result in exposure of their fields/crops to the attack of vertebrates. Damage to fields results in less production and hence less agricultural economic growth.

Several methods and techniques are used to reduce the damages caused by vertebrates in crops but nothing seemed to be satisfying. Different solutions are proposed for prevention of vertebrates' attack on crops [1]. Below are a few techniques which are already tested and still being researched in different parts of the world:

There are broadly two types of techniques that are being used:

- Non-Lethal: Below are a few Non-lethal techniques being used:
    - Guarding
    - Fencing / Netting
    - Chemical Repellents
    - Use of food decoys
- Lethal: Below are a few lethal techniques being used:
    - Shooting
    - Trapping and Poisoning

Experiments and tests have shown that non-lethal techniques are more effective than lethal techniques but still results are not substantial [1]. Different non-lethal techniques are effective for different types of crops and for different types of vertebrates but still no universal techniques are available that give satisfying results. Below are the major contributions of this article:

- Propose a low cost and power efficient prevention system to mitigate vertebrates crop raiding in precision agriculture
- Evaluate ZigBee performance in agricultural environment to ensure robust, reliable and efficient short-range communication
- Analyze factual multi-hop information transmission to realize low cost solutions

The rest of the paper is organized as follows. Section two shed light on a literature review about the usage of Wireless Sensor Network (WSN) in the agriculture to prevent damage of crops from vertebrates. Section three explains the model of the proposed system. The proposed system flow is discussed in section four. Section five describes the repelling and notifying component in the proposed system. Section six demonstrates the system break down. Section seven analyses the best arrangement of the wireless sensors in order to optimize their coverage on a given piece of land. The last section presents the conclusion and future enhancements.

## 2. LITERATURE REVIEW

Researchers throughout the world are working on different agricultural systems to reduce the damages of crops and increase the growth. Many pesticides have been invented that are intended for repelling and discouraging the pests from the crops [2]. All these researches are for a certain type of crops and there is no general method to prevent all the crops from damages from vertebrates.

According to an estimate, annual crop damages in Pakistan caused by vertebrates are more than 30 million rupees [3]. Similarly, Rodents inflict several losses to standing crops like wheat, paddy, sugarcane etc. According to an estimate the extent of damages from pests in Central Punjab alone is from 5 – 43% [3] which is alarming for a country whose economy depends hugely on Agriculture. In order to minimize these damages, prevention of crops from vertebrates is must. Less vertebrate attack means more crops and hence better economy.

Authors [4] have proposed a complete system architecture for WSN based greenhouse management. The author implements some sensors that monitors different greenhouse environment parameters like carbon dioxide level, water, humidity, temperature and light etc. This monitored data is then sent to the main application server using gateways and then after processing they decide to control the greenhouse parameters of the environment using pumps heaters and fans. The data is also available on a handheld device for the farmer. Similarly, authors [5] have also proposed relatively similar system named "Precision Agriculture Monitor System". The system monitors the agriculture parameters and then controls them. The system also incorporates WSN for data collection. They have implemented GPRS to send the data to the remote server. According to authors this system is also being used in some farms and the data provided by the system has been very useful for the farm management. Another similar system has been introduced by [6]. The proposed system monitors the soil and environment data on a farm and sends it to the server, from where it can eventually be sent to the farmer either using internet or a GSM network. All three of the systems however do not guarantee to prevent animals from entering the farm.

A lot of work is also being done to prevent the crops from birds most of those methods are the ones that have been described earlier or by the use of repellent spray. Stickley and Guarino [7] in their research have described that how a chemical named "Methiocarb" can repel the birds especially crows from sprouting corn. The authors have proved that the fields without spray resulted in 147% more damage than the fields sprayed with methiocarb. The research shows very significant results however the method is tedious, costly and time consuming.

A psychological Report [8] has been published by Pinel to use high-intensity, ultrasonic sound as a better rat trap. The sound waves produced can really reduce the rat invasion in the area of rat population, and also reduce the next generations of rat to colonize the un-inhabited area. Such as birds and animals run away by a gunshot or an explosion sound; the ultrasonic waves will just scare them rather than killing them. This method is non-invasive and environment friendly, as it does not chemically or ecologically interfere with the environment. The benefit of the proposed system by this paper is that it only turns on the sound once it has detected a presence of an animal or bird. It does not create continuous sound to pollute noise pollution for animals continuously.

Woo [9] presented an arrangement of scaring the animals entering the field through ultrasonic sounds. The system is claimed to be low cost because of its open source nature. Furthermore, the system is expected to be operated through smartphone.

Sendra et. al, [10] proposed a system that protects sheep and goats in the farms from the wolves. They monitored the heartbeats of the sheep and goats to analyze the danger around them as the heartbeats of these animals increase if another animal or a human is around them. Based on the several heartbeats' readings, signals are sent through a wireless sensor network to central station, that takes a decision of either sending a person in-charge or activates the visual or acoustic alarms to scare the predator.

Deshpande et. al [11] presented a model of protecting the fields from wild animals by using fence as a sensor and transmitting the presence of the wild animal through wireless sensor network to the central station. The system detects the entry of the wild animals and notifies the person in-charge through a SMS. Besides sending a SMS, the system also turns on the flash lights on the farm to highlight the presence of the animal in the field.

Most of the literature till date provides an insight into the various techniques proposed for detecting the presence of the animals in the field. Similarly, some literature provides an understanding of how farmers can be informed regarding the intrusion of animals in their farms using state-of-the-art systems. However, there is gap in literature when it comes to proposing a system that automatically repels the animals from the farms and informs the farmer about the intrusion of the animals. This paper presents a WSN based farm intrusion monitoring rather than greenhouse monitoring like many have worked before. Whenever it detects a motion precisely a biological motion within its range it only activates respective high intensity ultrasonic speakers to generate ultrasonic sound, and after some time it again stops. A human present nearby, will never know that the speaker is generating ultrasonic sounds. A notification to the farmer will also be sent so that he knows that the system is working and performing necessary actions.

## 3. PROPOSED SYSTEM MODEL

This research proposed a prevention system that helps farmers protect their fields from vertebrate attacks even if farmers are not available at their fields. The fundamental of this research is to develop a system which is efficient in performing the task and consumes low power for its operation, so that it is affordable for the farmers of a developing country. The proposed system is an application of WSNs and the article focuses on the coverage of WSNs by experimenting the communication range of short-range wireless technology, ZigBee. The information (presence of an animal in this case) is transmitted to neighboring nodes along with the measurement of appropriateness of the transfer time. Figure 1 shows the proposed system model for prevention of crops from vertebrates.

In this proposed system, motion sensors are placed around the field, which will detect any kind of motion within their range. These motion sensors are connected to a short range wireless technology module, ZigBee, through a circuit board which can transfer the information of motion detection from motion sensor to the central Repelling and Notifying System (RNS) wirelessly [12]. The approximate range of each sensor is around 10m [1][13].

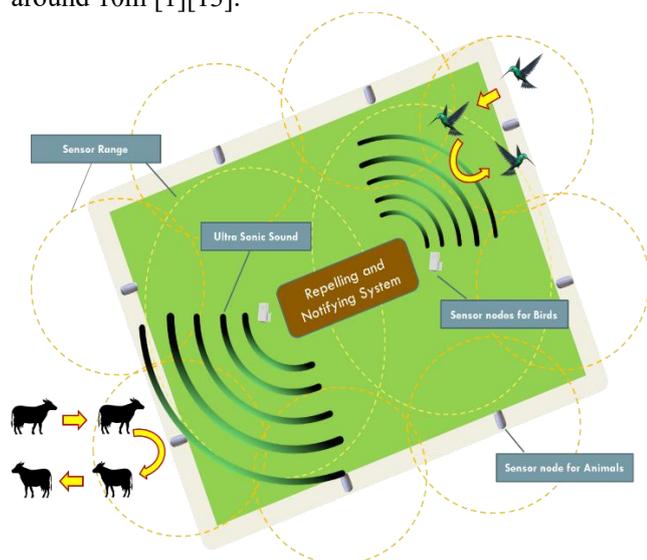

Fig. 1. Proposed System Model for prevention of crops from vertebrates

Whenever the sensor detects the nearby animal within its range, it sends this information to RNS through ZigBee communication which then activates the animal repeller device which generates ultrasonic sounds that are unbearable by the animals and as soon as animals hear that sound, they are forced to run away. For this research, the animals considered for proposing the system are, cows, horses, goats, sheep, dogs, cats and donkeys that are the most common animals found in agricultural areas of developing countries like Pakistan. Table 1 illustrates the approximate hearing ranges for the animals considered for this research.

Major reason of using ZigBee in this research is that ZigBee is low power and low cost short range wireless technology [14]. The amount of information to be transferred from sensor to RNS is few bytes which can be easily transferred using ZigBee [15]. Another reason of proposing the solution using wireless technology is to avoid the hassle of wired network in the field. Furthermore, as soon as the sensor detects the vertebrate, the information is sent to RNS. The information is also sent to the farmer using Short Messaging Service (SMS) through GPRS/GSM module used in RNS, so that the farmer gets alarmed by any intrusion of vertebrate in the field. Although farmer does not need to do anything as the proposed system will force the vertebrate to run away from the field, it is just an information system for the farmer.

Table 1. Approx. hearing ranges for the animals

| Animal | Approx. Range (Hz) |
|--------|--------------------|
| Cow    | 23 – 35,000        |
| Horse  | 55 – 33,500        |
| Sheep  | 100 – 30,000       |
| Dog    | 67 – 45,000        |
| Cat    | 45 – 64,000        |
| Goat   | 78 – 34,000        |
| Donkey | 10 – 40,000        |

## 4. PROPOSED SYSTEM FLOW

As soon as an animal or a bird is detected by the motion sensor, this information is sent to RNS through ZigBee (the ZigBee Pro modules have at least 800m range for clear line of sight) [19]. RNS after receiving this information generates the ultrasonic sounds from its repellers. As soon as the bird or the animal hears these unbearable sounds, they run away and the field is protected without farmer's availability. Fig. 2 below shows the proposed system flow. Furthermore, this information of animal/bird trying to enter the field is also sent to the farmer's phone by RNS which uses an Arduino + GPRS module to send the SMS.

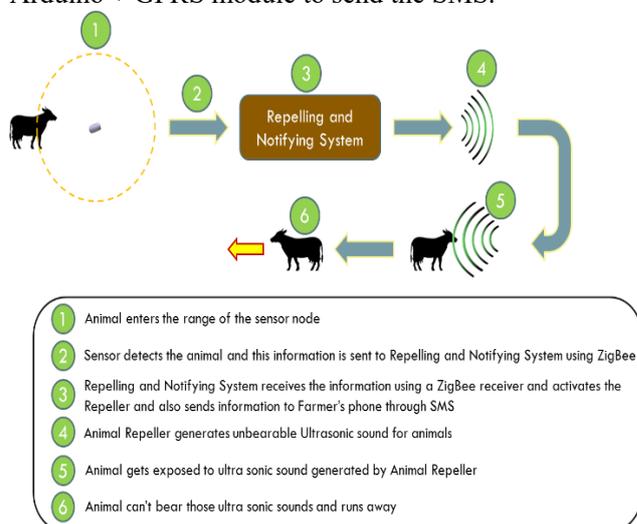

Fig. 2. Proposed System Life Cycle

## 5. REPELLING AND NOTIFYING SYSTEM

The central part of RNS is a microprocessor board, Arduino. Arduino is connected to a ZigBee module (Coordinator Device) which will receive information from its ZigBee end device attached to the motion sensor. Arduino is also connected to a GSM/GPRS module which is capable of sending SMS. An Animal Repeller is also connected to Arduino, which is normally in a sleep mode to save the energy, however they get activated by Arduino as soon as animal is detected by the motion sensor and information is transferred from motion sensor to Arduino board using ZigBee network. Animal repellers provide a range of 300 sq. meter, operate at 15KHz frequency and very low power of less than 200mA. A solar panel or a portable battery will be used to power up Arduino and animal repeller. Solar panel or a portable battery is selected to be used in this research because it might be difficult to arrange for electricity to power up these devices in remote fields, these portable batteries will provide flexibility in remote areas. Fig. 3 below shows the Repelling and Notifying System (RNS) for the proposed prevention system.

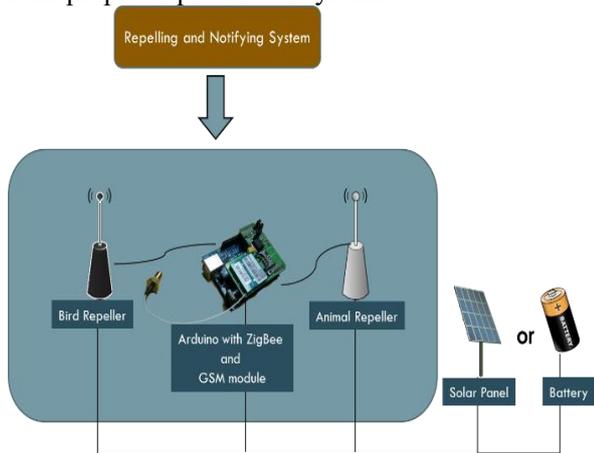

Fig. 3. Repelling and Notifying System for Proposed Prevention System

## 6. SYSTEM BREAKDOWN

Fig. 4 below shows the arrangement of the whole system. As shown in the figure below a motion sensor is connected to the ZigBee end device. The ZigBee end device will be powered through battery [16]. Motion sensor will get its power from ZigBee end device connected to it [17].

As illustrated in Fig. 4, whenever the motion sensor detects the motion within its range (around 10m), and this information from motion sensor is sent to the Arduino board through ZigBee network. Arduino after receiving this information wakes up the repellers which will generate ultrasonic sounds which will force animals to run away. At the same time, this information of intrusion of animal within the field is sent to the farmer on his mobile phone.

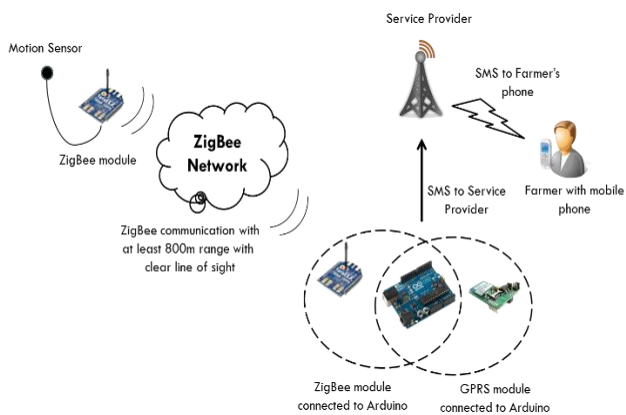

Fig. 4. System arrangement

Fig. 5 below shows a further breakdown of the system. An animal in sensor's vicinity of 10m is detected by the motion sensors which are covering the field. This information is sent to Arduino which activates the animal repeller.

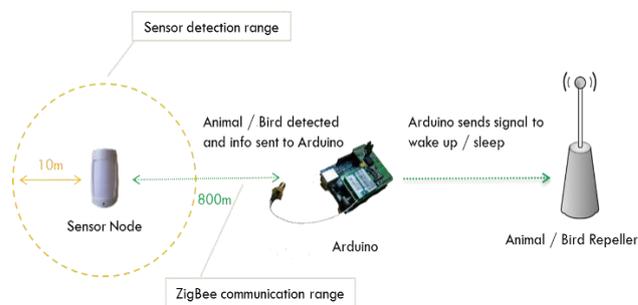

Fig. 5. System Breakdown (1)

The important feature to note over here is that the system is operating in a very low power mode until a motion is detected by the motion sensor. As soon as the motion is detected, the system wakes up and performs the necessary operations.

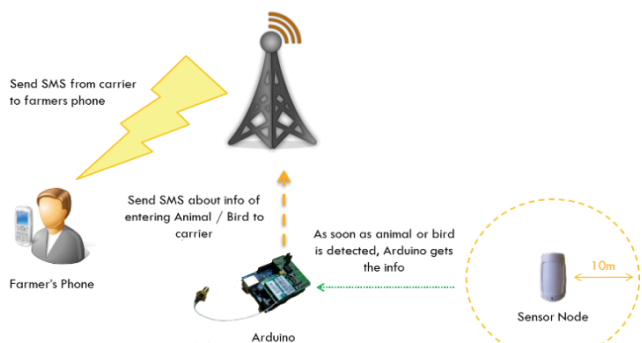

Fig, 6. System Breakdown (2)

Making system operate in a low power mode is the core of this research, as this system is targeted towards the developing countries, where providing power in a remote area (fields) is a challenge. For this purpose, throughout this system, portable power supplies e.g. Solar cells or Batteries are used. Major reason of using ZigBee network for transferring information in this system is because ZigBee is a low power and low cost technology. It needs very low power for its operation and also reduces the hassle of changing batteries since ZigBee modules can operate on same battery for years. [18]. Lee. et. al [19] provided an excellent comparative study of different short range wireless technologies evaluating their main features that include power consumption as well. The discussion provides an insight into how ZigBee devices can operate way longer than other wireless devices due to their very low power consumption.

Fig. 6 shows that as soon as the information of nearby animal/bird is sent to RNS, the Arduino with GSM module sends this information in the form of SMS to the carrier server which is then sent to the farmer's phone. In this way even farmer gets notified about any kind of animal/bird intrusion in the field. So this system is information rich and efficient in performing the task.

## 7. RESULTS AND DISCUSSIONS

Wireless communication is an important part of this research. It makes the flow of information from sensor to RNS wirelessly and makes the system less expensive and hassle free. It is very important to make sure that the system won't interrupt farmers' work or system won't get damaged while farmer is performing his daily farm activities like watering and seeding the field. If we use wired network for communication, the wires will be all around the field which will make the system less reliable and difficult to use. Wireless technologies and that too with low power are a perfect solution for information transfer in remote areas [18].

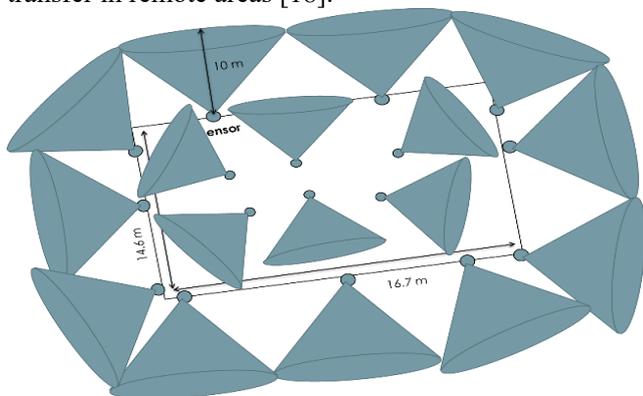

Fig. 7. Proposed arrangement of Sensors in the specific field

Fig. 7 shows the proposed arrangement of the motion sensors in the field. The typical range for a normal motion sensor is around 10m, so the sensors in the field are arranged in such a way that whole field is covered as shown in Fig 7. The arrangement of the sensors is directly proportional to the overall cost of the system, as with the increase in the area of field, the no. of sensors covering the area will also increase. As estimating the cost of the whole system depends upon the sensors required for the coverage of the area, Furthermore, scalability of the network is a broad problem which has to be covered taking into account several algorithms already available in the literature that is beyond the scope of this paper.

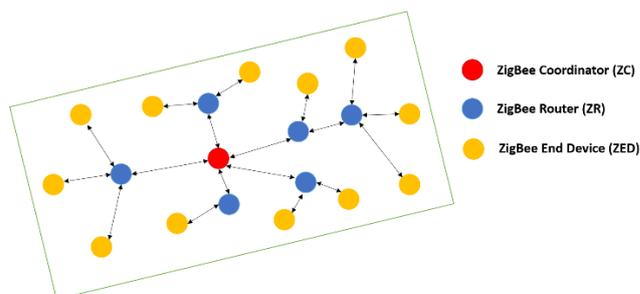

Fig. 8. Tree topology for sensor arrangement for proposed model

The proposed model is using ZigBee as a communication medium between the sensors, so the topologies compatible with ZigBee are considered for arrangement of sensors in the field. ZigBee supports Star, Tree and Mesh topology. Star topology is very basic and might not be feasible with large number of sensors in the field. Since the proposed model is designed for scalability hence increase in number of sensors might decrease the performance if star topology is used. Mesh topology is flexible with routing and provides good efficiency with large number of sensors, however, Mesh topology is very complex and nodes must perform computations for finding best route using the selected routing protocol which will increase the power consumption of the nodes and hence the whole system. For the proposed system, Tree topology is considered appropriate since it can handle large number of sensors and has less complexity. Fig. 8 shows the proposed Tree topology for the model under consideration.

Table 2. Simulation results for multi-hop ZigBee communication using Tree Topology

| Distance | Data (Bytes) | No. of Hops | Time (seconds) |
|---|---|---|---|
| 10 | 1 | 1 | 0.048 |
| | 50 | 1 | 0.090 |
| | 100 | 1 | 0.114 |
| 20 | 10 | 2 | 0.082 |
| | 50 | 2 | 0.120 |
| | 100 | 2 | 0.200 |
| 30 | 10 | 3 | 0.115 |
| | 50 | 3 | 0.224 |
| | 100 | 3 | 0.411 |

Several simulations were performed to gauge the efficiency of the tree topology for arrangement of sensors in the field. Simulations were performed on NS2 simulator with the arrangement shown in Fig. 8. For all the simulations, only 1 ZigBee Coordinator (ZC) was used along with maximum 2 ZigBee Routers (ZR) and 3 ZigBee End Devices (ZED). Each node, e.g., ZC, ZR and ZED are kept 10m from each other to keep simulation simple, however in real scenario, distance can change depending upon various factors [20].

Table 2 illustrates the simulation results with different set of distances (no. of hops) and amount of data sent from ZED and ZR to ZC. Table further illustrates the time taken by the nodes to send data to ZC. The data collected from the simulation was encouraging since, the time required for ZC to receive data is far less than the reaction time of the vertebrates entering the field. Furthermore, the amount of data, e.g., 10, 50 and 100 bytes seems to be sufficient for sending information about vertebrates entering the field, since the data will be small and precise without details of the vertebrate.

Fig. 9 below illustrates that with the increase in the distance and no. of hops of communication, receiving time of ZC increase. However, since the increase in time is very small, so it will not have a significant effect on the performance of the system.

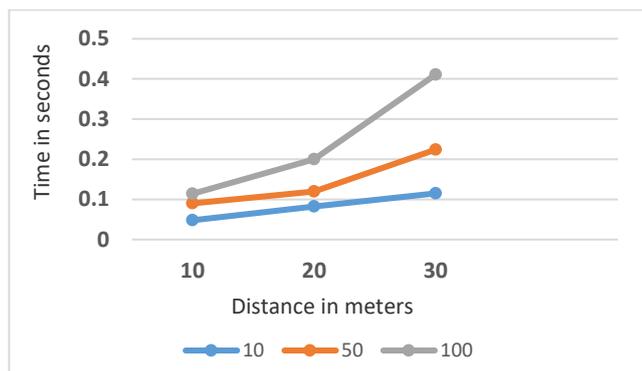

Fig. 9. Graph for multi-hop ZigBee communication using Tree Topology

Simulations show that tree topology works efficiently with the proposed model for arrangement of sensors, however, reliable data delivery must also be ensured to make this system robust [21]. For this purpose, several experiments were conducted to ensure reliability of data using Received Signal Strength Indicator (RSSI). Table 2 illustrates the strength of the received signal at ZC in order to receive data efficiently and reliability. RSSI value appeared to be between -62dBm to -86dBm which is still considered reasonable since the connection is intact and data was still delivered to ZC.

Table 3. Experimental results for multi-hop ZigBee communication through RSSI

| Distance | Data (Bytes) | No. of Hops | RSSI (dBm) |
|---|---|---|---|
| 10 | 1 | 1 | -67 |
|  | 50 | 1 | -68 |
|  | 100 | 1 | -62 |
| 20 | 10 | 2 | -77 |
|  | 50 | 2 | -79 |
|  | 100 | 2 | -75 |
| 30 | 10 | 3 | -82 |
|  | 50 | 3 | -85 |
|  | 100 | 3 | -86 |

As illustrated in Table 2 and Fig. 9, although ZC receives the information at a reasonable time which does not much effect the overall performance of the system. Similarly, Table II illustrates the transmission of data without any break in connection, however, network throughput can be a factor which further increases or decreases the efficiency and reliability of the system. Fig. 10 illustrates multi-hop ZigBee communication through RSSI based on data from Table 3.

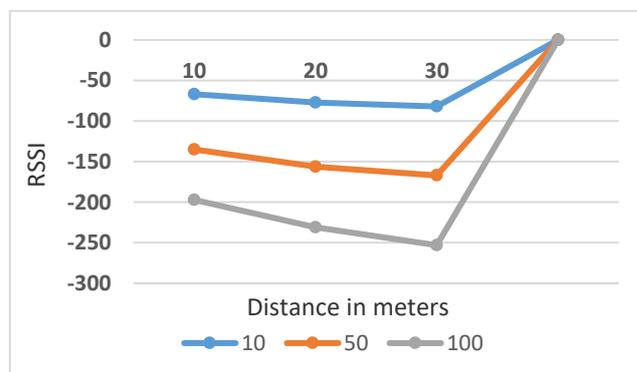

Fig. 10. Graph for multi-hop ZigBee communication through RSSI

In order to calculate throughput, several simulations were performed. Table 4 illustrates the results of the simulations for throughput.

Table 4. Simulation results for throughput in Tree Topology for ZigBee communication

| RSSI (dBm) | No. of Hops | Throughput (bits/sec) |
|---|---|---|
| < -50 | 1 | 40,000 |
| -50 | 1 | 35,000 |
| -60 | 1 | 28,000 |
| -70 | 1 | 25,000 |
| -80 | 1 | 20,000 |
| > -80 | 1 | 20,000 |

Table 4 illustrates the throughput for only single hop communication using Tree Topology for ZigBee communication. Throughout was analyzed based on RSSI value.

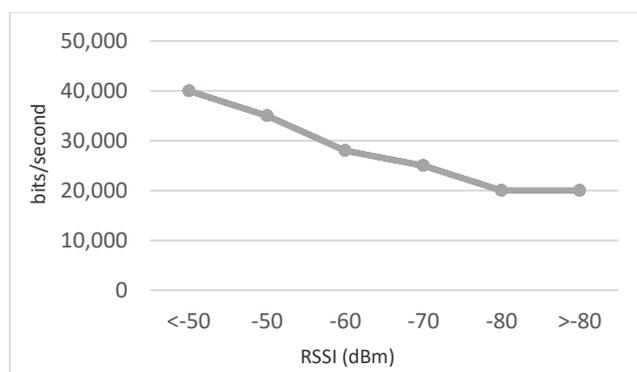

Fig. 11. Graph for throughput of ZigBee communication using Tree topology

As illustrated in Table 4, if RSSI value is less than -50dBm, there is maximum throughput which is around 40kbps. The reason being, the signal strength is maximum for RSSI values less than -50dBm. There can be many factors for this including, distance between the nodes, noise, obstacles, strength of sending device and capability of the receiving device. Similarly, throughout is good between -60dBm and -70dBm, however it starts reducing at -80dBm and below. Fig. 11 shows the graphical representation of the data presented in Table 3. Experimental and simulation results are encouraging and provides a foundation for developing farming system based on wireless sensor networks. Based on the results, it is evident that the proposed model provides efficient performance and reliable data delivery using ZigBee communication for arrangement of sensors using Tree topology. The proposed model can set a foundation for developing other similar systems for farming, e.g., crop growth monitoring, water level monitoring and alarming, sanity level warning and farms intrusion detection etc.

## 8. CONCLUSION

The paper presented a prevention system for the farmers that can help them protect their crops from attack of different vertebrates. The prevention of crops will result in more crop growth and hence better economy. In order to keep the cost of the system low, ZigBee is used which is low power and low cost technology. Combination of ZigBee, Arduino and Repelling systems enable this system to detect animals in the field using motion sensors and generate ultrasonic sounds which are unbearable for animals and make them run away from the field.

Furthermore, the paper brings an idea of notifying the farmer about intrusion of vertebrate into the field through SMS which will open a new horizon of opportunities for other research projects. This research will lay a foundation for a notifying system through SMS for the farmers that can be used in different researches, for example farmers can be notified about ready to harvest crops, time to water the crops and many more, the possibilities are endless.

Lastly, simulation and experimental data helps in proving the efficiency and reliability for the arrangement of the sensors using tree topology. Timely data delivery, efficient multi-hop communication, reliable data transmission and low cost of the system enables this system to be implemented in developing countries that are heavily dependent on agriculture for improving their socio-economic growth.